\documentclass[english,aps,prb,twocolumn,superscriptaddress]{revtex4}
\usepackage{amsmath}
\usepackage{amssymb}
\usepackage[dvipdfmx]{graphicx}
\usepackage[dvipdfmx]{color}
\usepackage{ulem}

\newcommand{\bvec}[1]{\mbox{\boldmath $#1$}}

\makeatletter
\usepackage[colorlinks=true,urlcolor=blue,citecolor=blue,linkcolor=blue,breaklinks=true,dvipdfmx]{hyperref}
\makeatother
\begin{document}

\title{Interplay of Pomeranchuk instability and superconductivity\\ 
in the two-dimensional repulsive Hubbard model}

\author{Motoharu Kitatani}
\affiliation{Department of Physics, University of Tokyo, Hongo, Tokyo 113-0033, Japan}

\author{Naoto Tsuji}
\affiliation{RIKEN Center for Emergent Matter Science (CEMS), Wako 351-0198, Japan}

\author{Hideo Aoki}
\affiliation{Department of Physics, University of Tokyo, Hongo, Tokyo 113-0033, Japan}
\affiliation{Electronics and Photonics Research Institute,
Advanced Industrial Science and Technology (AIST), Tsukuba, Ibaraki 305-8568, Japan}

\date{\today}
\begin{abstract}
Interplay of Pomeranchuk instability (spontaneous symmetry breaking of the Fermi surface) and $d$-wave superconductivity is studied for 
the repulsive Hubbard model 
on the square lattice with the dynamical mean field theory
combined with the fluctuation exchange approximation (FLEX+DMFT).  
We show that the four-fold symmetric Fermi surface becomes unstable against a spontaneous distortion into two-fold near the van Hove filling,
where the symmetry of superconductivity coexisting with the Pomeranchuk-distorted Fermi 
surface is modified from the $d$-wave pairing to $(d+s)$-wave.  
By systematically shifting the position of van Hove filling 
with varied second- and third-neighbor hoppings, 
we find that 
the transition temperature $T_{\rm c}^{\rm PI}$ for the Pomeranchuk instability
is more sensitively affected by the position of van Hove filling than the superconducting $T_{\rm c}^{\rm SC}$. 
This implies that the filling region for strong Pomeranchuk instability and 
that for the $T_{\rm c}^{\rm SC}$ dome 
can be separated, and that Pomeranchuk instability can appear
even if the peak of $T_{\rm c}^{\rm PI}$ is lower than the peak of $T_{\rm c}^{\rm SC}$.
An interesting finding is that 
the Fermi surface distortion can enhance the superconducting 
$T_{\rm c}^{\rm SC}$ in the overdoped regime, which is 
explained with a perturbational picture for small distortions. 
\end{abstract}



\maketitle

\section{Introduction}
High-Tc cuprate superconductors harbor many fundamental questions, which 
challenge elaborate numerical analysis on 
superconductivity, magnetism and other properties. 
Specifically, there is growing realization that 
various instabilities can exist along with superconductivity \cite{cuprate-review}, and the relation between various charge instabilities 
and superconductivity in the cuprates is now being intensively studied \cite{YBCO-charge-order,Bi-charge-order1,Bi-charge-order2,Hg-charge-order}.  
Also, some experiments suggest a spontaneous breakdown of the 
four-fold symmetry of electronic states in the tetragonally 
structured cuprates, which is viewed as a kind of 
``electronic nematicity"\cite{YBCO-transport,YBCO-neutron,BiSCCO-STS}.  
There are some explanations for the nematicity, e.g. in the context of fluctuating stripe orders \cite{fluctuating-stripe}.  
Pomeranchuk instability, a spontaneous breaking of four-fold symmetry of the Fermi surface without lattice distortion,
is evoked as another possible candidate for nematicity in cuprate superconductors \cite{tJ-PI-cuprate}.

The presence of Pomeranchuk instability in two-dimensional lattice models has been suggested in Refs.~\onlinecite{PI-Metzner,PI-Yamase},
where the forward scattering was found to develop to induce Pomeranchuk instability.  
Subsequently, properties of this instability were studied primarily in mean-field models (``f-model"), 
where the electrons interact only via forward scattering \cite{f-model-mean-field,f-model-fRG}.
For the two-dimensional (2D) Hubbard model on the square lattice,
a representative model for cuprates,
the existence of this instability is yet to be fully 
clarified microscopically. 
Functional renormalization group (fRG) calculations suggest that
the superconducting fluctuation is stronger than Pomeranchuk instability \cite{fRG-temp_cut},
while other numerical renormalization-group approaches suggest Pomeranchuk instability to be stronger around van Hove fillings\cite{PI-flow-equation}.
Gutzwiller wave functions combined with an efficient diagrammatic expansion technique (DE-GWF) 
obtained a ground state with a coexistence of the nematic order and superconductivity in the 2D Hubbard model \cite{nematic-dSC-DE-GWF},
which is also observed with the renormalized perturbation theory for the weak-coupling case \cite{renormalized-PT}.
Also, the dynamical cluster approximation (DCA) and cellular dynamical-mean-field theory (CDMFT)
showed large responses against small distortions of the lattice\cite{DCA-nematic,DCA-CDMFT-nematic}, 
from which a possibility of spontaneous symmetry breaking is suggested to occur at lower temperatures or for larger cluster sizes.
While these results suggest that the 2D repulsive Hubbard model has 
a strong tendency toward the Pomeranchuk instability, 
whether or not this instability has higher transition temperature ($T_{\rm c}^{\rm PI}$)
than that of superconductivity ($T_{\rm c}^{\rm SC}$) has yet to be elaborated. 
More importantly, the relation between the Pomeranchuk instability and superconductivity (e.g., whether they are cooperative or competing) 
is an intriguing question. 
While a mean-field study for a phenomenological model suggests that 
they are competing with each other with 
$T_{\rm c}^{\rm SC}$ suppressed
in the coexistence region \cite{dSC-dPI-mean-field}, 
the relation should be clarified by going beyond mean-field approaches.

Given the situation, we study in the present paper 
superconductivity and Pomeranchuk instability in the 
intermediate correlation regime by evoking FLEX+DMFT\cite{MBPT+DMFT,FLEX+DMFT}, a 
diagrammatic extension of the dynamical mean field theory (DMFT) \cite{d-infinite,DMFT,DMFT_review},
which takes account of the spin and charge fluctuation effects 
on top of the DMFT local self-energy, and can reproduce the dome 
structure in $T_{\rm c}^{\rm SC}$ \cite{FLEX+DMFT}. 
The advantages of this method are first, 
there is no finite-size effects unlike in DCA and CDMFT, which should be important for capturing 
small Fermi surface deformations,  
and second, we can calculate finite-temperature regions in contrast to DE-GWF
to capture 
the effect of this nematicity on the superconducting $T_{\rm c}^{\rm SC}$.
This also enables us to systematically examine the relation between 
superconductivity and Pomeranchuk instability when 
the electron band filling and the second and further neighbor hoppings 
($t^{\prime}, t^{\prime\prime}$) are varied.  
After confirming the existence of 
Pomeranchuk instability around the van Hove filling consistently with 
the previous works, we shall study the superconducting phase, which 
reveals that the symmetry of the gap function is changed from 
the ordinary $d$-wave pairing to ($d+s$)-wave \cite{DCA-nematic,nematic-dSC-DE-GWF}.  
Interestingly, the Fermi surface distortion 
can enhance the superconductivity in the overdoped (or strongly frustrated) regime with larger $t^{\prime}, t^{\prime\prime}$.  
We shall explain this $T_{\rm c}^{\rm SC}$ enhancement with a perturbation picture 
for small Fermi-surface distortions, and 
also with the random phase approximation (RPA) in the weak-coupling regime.

Another finding here is that 
the Pomeranchuk instability temperature $T_{\rm c}^{\rm PI}$ is more sensitive to $(t^{\prime}, t^{\prime\prime})$, hence 
the Fermi surface warping, than the superconducting $T_{\rm c}^{\rm SC}$.
This contrasts with 
the previous mean-field calculations \cite{dSC-dPI-mean-field} that showed almost the same filling dependence for the two 
transition temperatures.  This should come from the fact that the present formalism 
takes account of the filling dependence of the pairing interaction 
beyond mean-field levels.  
The result also implies that the superconducting $T_{\rm c}$ dome 
and that for Pomeranchuk instability can be separated.

\section{Formulation}
We consider the standard repulsive Hubbard model on the  square lattice 
with a Hamiltonian, 
\begin{equation}
	{\cal H} = \sum_{{\boldsymbol{k}},\sigma} \epsilon({\boldsymbol{k}})c_{{\boldsymbol{k}},\sigma}^{\dag}c_{{\boldsymbol{k}},\sigma} + U \sum_{i} n_{i,\uparrow} n_{i,\downarrow},
\end{equation}
where $c_{{\boldsymbol{k}},\sigma}^{\dag}$ creates 
an electron with wave-vector ${\bvec k} = (k_{x},k_{y})$ and spin $\sigma$, 
$U$ is the on-site Coulomb repulsion, 
and $n_{i,\sigma} = c_{i,\sigma}^{\dag} c_{i,\sigma}$. 
In the presence of second-neighbor ($t^{\prime}$) and 
third-neighbor ($t^{\prime \prime}$) hopping parameters, 
the 2D band dispersion is given as
\begin{align}
	\epsilon&({\boldsymbol k}) = -2t({\rm cos}\; k_x + {\rm cos}\; k_y) \notag \\
							&\quad - 4t^{\prime} {\rm cos}\; k_x \;{\rm cos}\; k_y -2t^{\prime \prime}({\rm cos}\; 2k_x + {\rm cos}\; 2k_y)- \mu,
\label{dispersion}
\end{align}
where $t$ is the nearest-neighbor hopping (the unit of energy hereafter), 
and $ \mu $ the chemical potential.  
We basically adopt $ t^{\prime}=-0.20t $, $ t^{\prime\prime}=0.16t$, 
which are determined to fit the band calculation for a typical hole-doped single-layer cuprate, 
HgBa$_2$CuO$_{4+\delta}$ \cite{nishiguchi_prb,sakakibara}.

For the numerical procedure, we employ FLEX+DMFT method, which is a kind of diagrammatic extension of DMFT, 
where the fluctuation exchange approximation (FLEX) \cite{FLEX} 
and the DMFT are combined with a double self-consistency loop.
This kind of scheme has been considered in Refs.\onlinecite{QMC-FLEX,MBPT+DMFT}, and 
has recently been formulated through Luttinger-Ward functional 
with applications to superconducting states in Ref.\onlinecite{FLEX+DMFT}.
The latter can describe a $T_{\rm c}^{\rm SC}$ dome against the band filling 
along with a 
spectral weight transfer.  These are the virtue of 
FLEX+DMFT that corrects the overestimated local-FLEX self-energy in a 
filling-dependent manner. 
In FLEX+DMFT, the self-energy is calculated through the FLEX self-energy and DMFT self-energy ($\Sigma_{\rm imp}$) as
\begin{equation}
	\Sigma(k) = \Sigma_{\rm FLEX}(k) - \Sigma^{\rm loc}_{\rm FLEX}(k) + \Sigma_{\rm imp}(\omega_n),
\end{equation}
where the FLEX self-energy $\Sigma_{\rm FLEX}(k)$ is given as
\begin{align}
&
	\Sigma_{\rm FLEX}(k) = 
	\frac{1}{N_{\bm{k}} \beta} \sum_{k^{\prime}} \Bigl[
	\frac{3}{2}U^2 \frac{\chi_{0}(k-k^{\prime})}{1-U \chi_{0}(k-k^{\prime})} \notag \\
&
	\quad+ \frac{1}{2}U^2 \frac{\chi_{0}(k-k^{\prime})}{1+U\chi_{0}(k-k^{\prime})}
	- U^2\chi_{0}(k-k^{\prime}) \Bigr] G(k^{\prime}).
\label{FLEX-selfenergy}
\end{align}
Here $N_{\bm{k}}$ and $ \beta $ are the total number of $k$-points and inverse temperature, respectively,
$k \equiv (\omega_{n},{\boldsymbol k})$ with $\omega_{n}$ the Matsubara frequency for fermions, $G(k)$ Green's function, and
\begin{equation}
	\chi_{0}(q) = -\frac{1}{N_{\bm{k}} \beta}\sum_k G(k+q)G(k)
\label{chi}
\end{equation}
is the irreducible susceptibility.
The local part of the FLEX self-energy, $\Sigma^{\rm loc}_{\rm FLEX}$, 
is computed by replacing  Green's function $G$ with the local one, 
$G_{\rm loc} \equiv (1/N_k) \sum_{\boldsymbol k}G(k)$, 
in Eqs.~(\ref {FLEX-selfenergy}) and (\ref {chi}).

For calculating the DMFT self-energy, $\Sigma_{\rm imp}$, 
we need to solve the impurity problem in DMFT.  
Here we employ the modified iterative perturbation theory (modified IPT) as 
the impurity solver.
In this method, the original IPT is
modified for the systems having no particle-hole symmetry \cite{mIPT}, thus 
applicable to frustrated or non-half-filled cases. 
This is not computationally expensive,
which enables us to scan over various parameter regions.
We have checked, by using ALPS library\cite{ALPS1,ALPS2}, that the continuous-time quantum Monte 
Carlo (CT-QMC) impurity solver\cite{CTQMC,CTQMC-review}
gives similar results even away from half-filling in the intermediate-coupling regime [see Fig.~\ref{fig:eta_U4}(a)].

After obtaining Green's function, we plug it into 
the linearized Eliashberg equation, 
\begin{equation}
	\lambda \Delta(k) = -\frac{1}{N_{\bm{k}}\beta}\sum_{k^{\prime}} V_{\rm eff}(k-k^{\prime})|G(k^{\prime})|^2\Delta(k^{\prime}),
\label{eliashberg}
\end{equation}
where $\Delta(k)$ is the anomalous self-energy, while
\begin{equation}
	V_{\rm eff}(k) = U + \frac{3}{2}U^2 \frac{\chi_0(k)}{1-U\chi_0(k)} - \frac{1}{2}U^2 \frac{\chi_0(k)}{1+U\chi_0(k)}
\end{equation}
is the effective pairing interaction, and 
$\lambda$ the eigenvalue of Eliashberg's equation. Superconducting $T_{\rm c}^{\rm SC}$ is determined as 
the temperature at which $\lambda = 1$.
In the right-hand side of Eq.~(\ref {eliashberg}), we have neglected the local DMFT vertex contribution,
whose validity is discussed in Appendix \ref{appendix}.

To allow the Pomeranchuk instability to occur, we introduce a seed to deform
the Fermi surface in the initial input 
for the self-energy as
$\Sigma_{\rm initial} = 0.05t({\rm cos}\; k_x - {\rm cos}\; k_y)$.
While we linearize the anomalous part of Green's function as being infinitesimal, we can deal with finite Pomeranchuk order parameters, so that 
we can discuss superconductivity for finite distortions in this formalism.

\section{Results} \label{sec:Results}
\subsection{Pomeranchuk instability}

\begin{figure}
\begin{centering}
\includegraphics[width=1\columnwidth]{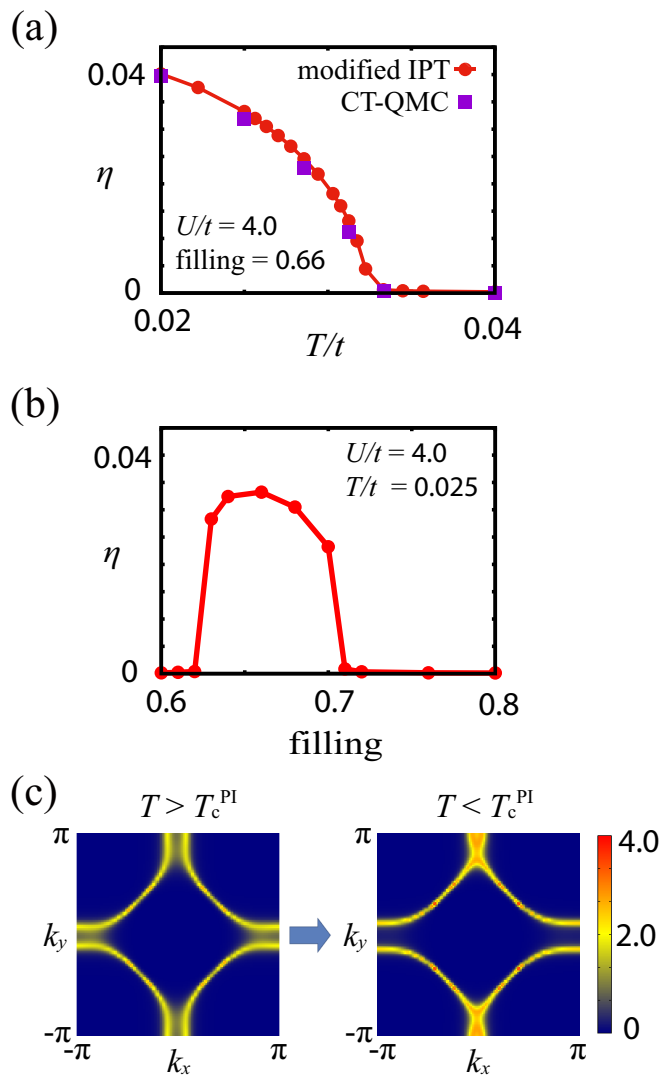}
\par\end{centering}
\caption{(Color online) (a) Temperature dependence and (b) filling dependence of the Pomeranchuk order parameter $\eta$ for $U/t=4.0,(t^{\prime}, t^{\prime \prime}) = (-0.20, 0.16)$.
In (a), the circles (squares) represent the results of FLEX+DMFT
with the modified IPT (CT-QMC) as a DMFT impurity solver.
(c) Fermi surface [as represented by the color-coded spectral weight $A({\boldsymbol k},\omega=0)$]
with $n=0.66,U/t=4.0,(t^{\prime}, t^{\prime \prime}) = (-0.20, 0.16)$, for $T=0.0333t>T_{\rm c}^{\rm PI}$ ($\beta t=30$; left) and $T=0.0286t<T_{\rm c}^{\rm PI}$ ($\beta t=35$; right).
}
\label{fig:eta_U4}
\end{figure}

\begin{figure}
\begin{centering}
\includegraphics[width=1\columnwidth]{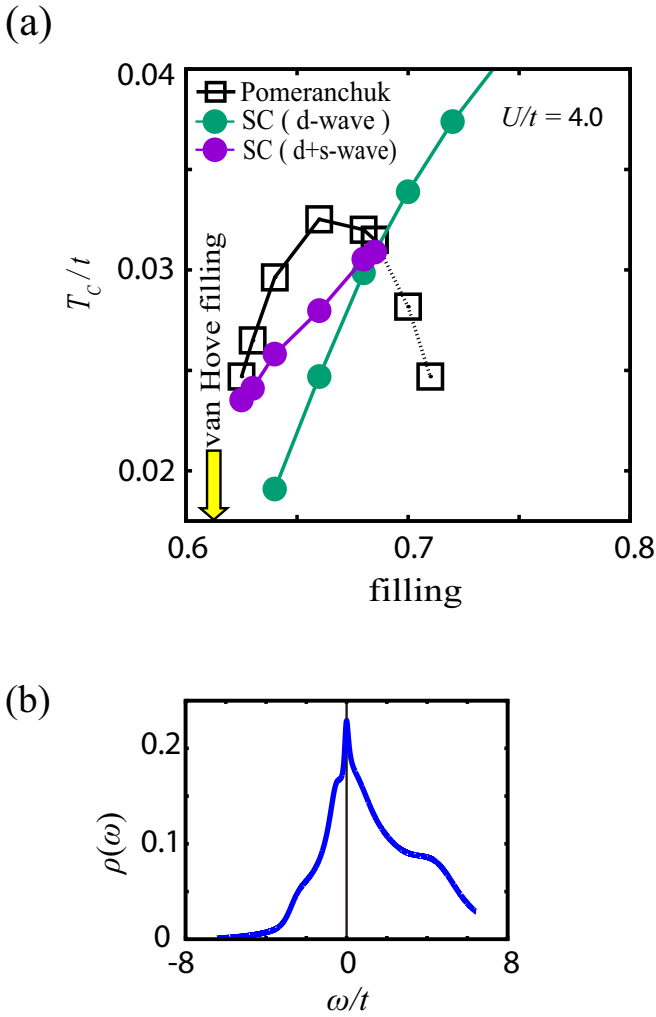}
\par\end{centering}

\caption{(Color online)
(a) Phase diagram against temperature $T/t$ and band filling $n$ for $U/t=4.0,(t^{\prime}, t^{\prime \prime}) = (-0.20, 0.16)$. Shown are the superconducting $T_{\rm c}^{\rm SC}$
with undistorted Fermi surface (green circles), 
superconducting $T_{\rm c}^{\rm SC}$ with Fermi surface distortion (purple circles), and Pomeranchuk $T_{\rm c}^{\rm PI}$ (black squares).  
The dotted line represents $T_{\rm c}^{\rm PI}$ when we ignore the superconductivity.
The yellow arrow indicates the van Hove filling in the interacting system. 
(b) Density of states at the filling indicated by the yellow arrow in (a) 
for $\beta t=20,U/t=4.0,(t^{\prime}, t^{\prime \prime}) = (-0.20, 0.16)$.}
\label{fig:phase_eta_U4}
\end{figure}

The Pomeranchuk order parameter $\eta$ can be defined,
for the originally four-fold cosine bands, as
\begin{equation}
	\eta = \sum_{\boldsymbol k} ({\rm cos}\;k_{y} - {\rm cos}\;k_{x}) \left< c^{\dag}_{\boldsymbol k}c_{\boldsymbol k} \right>,
\end{equation}
and we display the result against temperature and band filling in Fig.~\ref{fig:eta_U4}(a),(b), respectively.  
We can see that the order parameter starts to grow continuously 
with decreasing temperature, which indicates a second-order phase transition. 
If we turn to the filling dependence, we observe the order parameter abruptly 
grows around the edges of the Pomeranchuk phase, 
indicative of transferring to a first-order phase transition consistently with the previous work \cite{f-model-mean-field}.
Hereafter, we focus on the filling region around the peak of 
$T_{\rm c}^{\rm PI}$, where the transition is of second order.

If we look at the Fermi surface in Fig.~\ref{fig:eta_U4}(c) 
for $U/t=4.0$ and $n=0.66$, we can see that the 
Fermi surface, identified as the ridges in the spectral function $A({\boldsymbol k},\omega=0)$
obtained with the Pad\'{e} approximation, 
indeed becomes distorted at lower temperatures, $T<T_{\rm c}^{\rm PI}$.

The phase diagram is displayed in Fig.~\ref{fig:phase_eta_U4}(a), where 
we can see that the Pomeranchuk instability temperature 
$T_{\rm c}^{\rm PI}$, which is determined as the 
temperature at which $\eta$ becomes nonzero, 
is peaked 
around $n=0.66$ for the present parameter set
($U/t=4.0,(t^{\prime}, t^{\prime \prime}) = (-0.20, 0.16)$). 
An yellow arrow indicates the van Hove 
filling in the interacting system at which the spectrum is peaked at the Fermi energy, 
see Fig.~\ref{fig:phase_eta_U4}(b), in which we have obtained the
density of states 
with the Pad\'{e} approximation 
and confirmed the peak position does not change for $T>T_{\rm c}^{\rm PI}$.  
The fact that the Pomeranchuk instability tends to be 
strong near this filling is consistent with the previous results \cite{f-model-mean-field,dSC-dPI-mean-field}.
The peak in the Pomeranchuk dome does not precisely coincide with the van Hove filling, 
which may be an effect of the asymmetric density of states \cite{f-model-mean-field} as in Fig.~\ref{fig:phase_eta_U4}(b). 
By contrast, the superconducting $T_{\rm c}^{\rm SC}$ in 
the present result is a monotonically decreasing 
function of the hole doping around the van Hove filling.

\begin{figure}
\begin{centering}
\includegraphics[width=1\columnwidth]{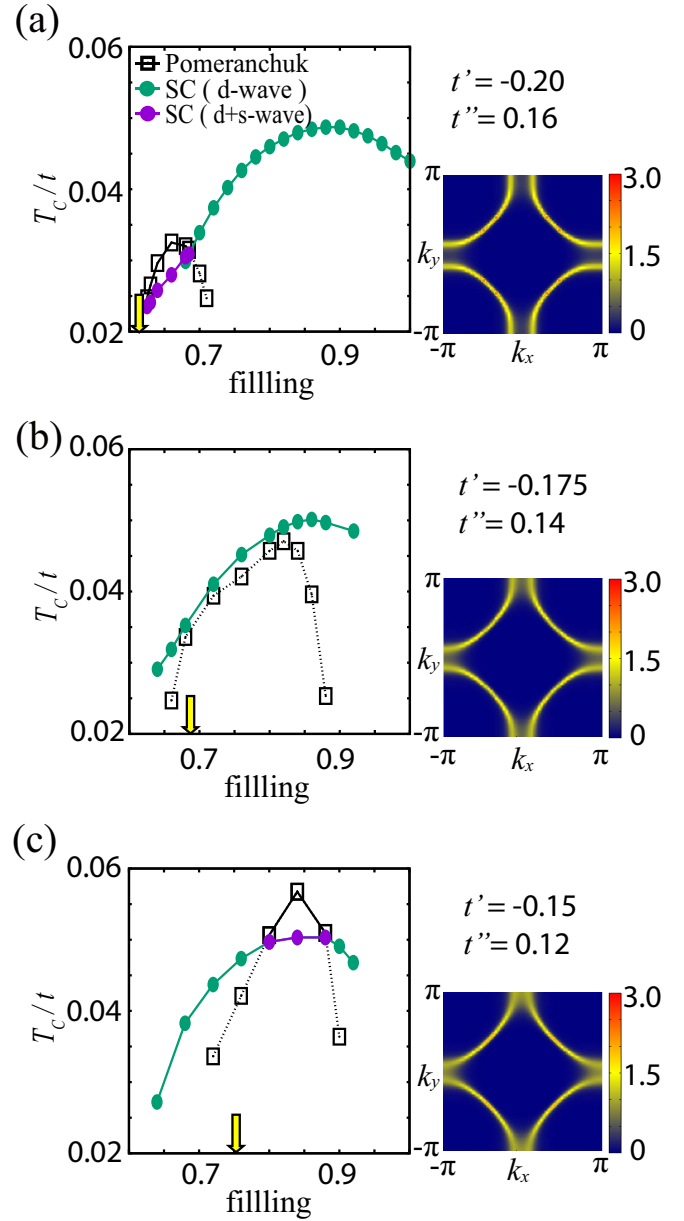}
\par\end{centering}
\caption{(Color online) 
The superconducting and Pomeranchuk phase boundaries for $U/t=4.0$ (left panels)
and the spectral weight $A({\boldsymbol k},\omega=0)$ for $n=0.80, \beta t =20, U/t=4.0$ (right)
are shown for $(t^{\prime}, t^{\prime \prime}) = (-0.20, 0.16)$ (a), 
$(-0.175, 0.14)$ (b), 
and $(-0.15, 0.12)$ (c).  
The symbols are the same as in Fig.~\ref{fig:phase_eta_U4}(a), 
and yellow arrows indicate respective van Hove fillings in the interacting system.
}
\label{fig:tp-dependence}
\end{figure}

This contrasts with the previous mean-field calculations \cite{dSC-dPI-mean-field} which ignore the filling dependence of the 
effective pairing interaction, where 
both $T_{\rm c}^{\rm SC}$ and $T_{\rm c}^{\rm PI}$ are peaked around the van Hove filling.  
Thus the present result indicates that the filling dependence of the effective interaction has an important effect of rendering 
distinction of optimal doping levels between the Pomeranchuk $T_{\rm c}^{\rm PI}$ dome and the superconducting $T_{\rm c}^{\rm SC}$ dome. 
To confirm this, let us systematically 
vary the second- and third-neighbor 
hopping parameters $(t^{\prime},t^{\prime \prime})$ in Fig.~\ref{fig:tp-dependence}, which change the Fermi surface warping as well as the van Hove filling.
We can see that, for a fixed $n=0.80$, the change in the parameters
shifts the distance of the filling from the van Hove filling
as represented by the blurring of the spectral function around $(0,\pi),(\pi,0)$.
Left panels in Fig.~\ref{fig:tp-dependence} plot the phase diagrams for 
three typical cases with different Fermi surface warping.  
We find that the Pomeranchuk $T_{\rm c}^{\rm PI}$ drastically changes 
along with the van Hove filling (yellow arrows), 
while the superconducting $T_{\rm c}^{\rm SC}$ is much less sensitive.

We can thus conclude that, despite both of superconductivity 
and Pomeranchuk instability 
being Fermi surface instabilities affected by the 
spectral weight at the Fermi energy, 
the Pomeranchuk instability is much more sensitive to the Fermi surface shape 
(distance from the van Hove filling).  
This implies that we can separate the dominant regions for 
the two instabilities by changing
the position against the van Hove filling (dominated by $t^{\prime},t^{\prime \prime}$).

\subsection{Superconductivity under Fermi surface distortions}
\begin{figure}
\begin{centering}
\includegraphics[width=1\columnwidth]{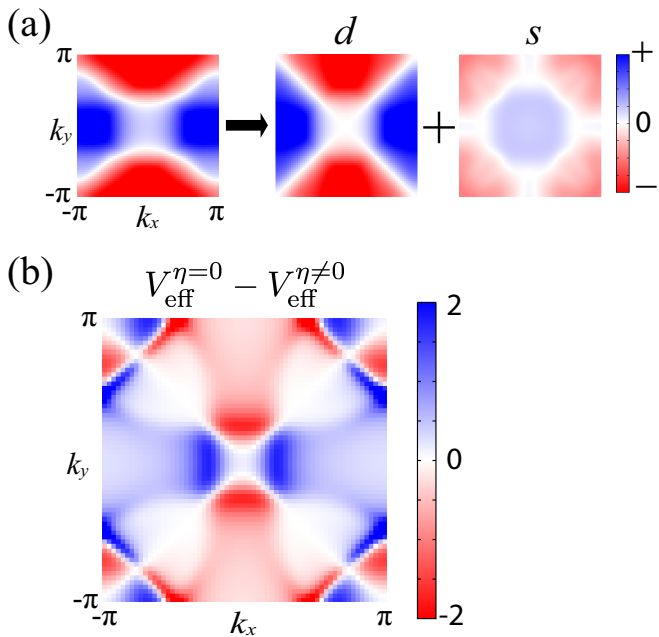}
\par\end{centering}
\caption{(Color online)
(a) Momentum dependence of the gap function for $T=0.0286t<T_c^{\rm PI}$ 
with  $n=0.66,U/t=4.0,(t^{\prime}, t^{\prime \prime}) = (-0.20, 0.16)$ (left panel), which can be 
decomposed into a $d$-wave part and an (extended) $s$-wave (four-fold symmetric) part (right). 
(b) Difference in the pairing interaction with the Fermi surface distortion ($V_{\rm eff}^{\eta \neq 0}$) and without ($V_{\rm eff}^{\eta = 0}$), 
for $n=0.66,\beta t =31, U/t=4.0,(t^{\prime}, t^{\prime \prime}) = (-0.20, 0.16)$.
}
\label{fig:phase_compare}
\end{figure}

Now, an intriguing issue is how superconductivity behaves in the 
presence of the Pomeranchuk Fermi-surface distortion.  
If we look at the superconducting order parameter 
in Fig.~\ref{fig:phase_compare}(a), the pairing symmetry 
is seen to be distorted from the ordinary $d$-wave 
to $d+$(extended)$s$-wave.  
Here, an interesting observation is that the superconducting $T_{\rm c}^{\rm SC}$ can 
be {\it enhanced} by the Pomeranchuk distortion of the Fermi surface.
Indeed, if we go back to Fig.~\ref{fig:phase_eta_U4}, we have also plotted 
the superconducting $T_{\rm c}^{\rm SC}$ (green dots) 
when the four-fold Fermi surface is 
artificially imposed below Pomeranchuk $T_{\rm c}^{\rm PI}$.  
We can see the $T_{\rm c}^{\rm SC}$ with the distorted Fermi surface (purple dots) is actually higher.

To identify the origin of this enhancement, we can compare 
the pairing interaction between the cases of 
Pomeranchuk-distorted and the four-fold-imposed Fermi surfaces. 
Figure~\ref{fig:phase_compare}(b) plots the difference of the two cases 
for the same parameters ($U/t=4.0$, $n=0.66$ and $\beta t =31$).  
We can see that the Pomeranchuk instability distorts the 
pairing interaction, where the difference has 
a $d$-wave-like sign reversal.

To pin-point the origin of the distortion effect on the superconducting $T_{\rm c}^{\rm SC}$,
we can consider the perturbational effect for 
small distortions, based on a general linearized gap equation,
\begin{equation}
	\lambda \phi(k) = -\frac{1}{N_{\bm{k}}\beta}\sum_{k^{\prime}} K(k,k^{\prime}) \phi(k^{\prime}), 
\label{general-eliashberg}
\end{equation}
where $\phi(k)=|G(k)| \Delta (k)$, while $K(k,k^{\prime})$ is the kernel, given 
as $K(k,k^{\prime})=|G(k)|V_{\rm eff}(k-k^{\prime})|G(k^{\prime})|$ in FLEX+DMFT
(as seen by multiplying $|G|$ to both sides of Eq.~(\ref {eliashberg})). 
If we consider small $d$-wave-like distortions [as displayed in Fig.~\ref{fig:phase_compare}(b)] for this kernel,
\begin{equation}
	K(k,k^{\prime}) \rightarrow K(k,k^{\prime}) + \delta K^{\rm d}(k,k^{\prime}),
\end{equation}
the first-order perturbation for the maximum eigenvalue $\lambda_{\rm max}$ 
satisfies
\begin{equation}
	\delta \lambda_{\rm max}^{\rm (1)} = \sum_{k,k^{\prime}} \phi^{*}_{\rm max}(k) \delta K^{\rm d}(k,k^{\prime}) \phi_{\rm max}(k^{\prime}) = 0,
\end{equation}
where $\phi_{\rm max}$ is the eigenvector for $\lambda_{\rm max}$.  
Namely, $\delta \lambda_{\rm max}^{\rm (1)}$ identically vanishes due to the $d$-wave nature of the $\delta K^{\rm d}$, 
so that the leading term is the second-order one,
\begin{equation}
	\delta \lambda_{\rm max}^{\rm (2)} = \sum_{i,k,k^{\prime}} \frac{|\phi^{*}_{\rm max}(k) \delta K^{\rm d}(k,k^{\prime})\phi_{i}(k^{\prime})|^2 }{\lambda_{\rm max}-\lambda_{i}}
	> 0,
\end{equation}
where $i$ is the index for the eigenvalue $\lambda_{i}$ and eigenvector $\phi_{i}$ of the kernel matrix $K$.
Since this expression is positive-definite, small $d$-wave deformations of the kernel in the linearized gap equation always enhance the superconducting $T_{\rm c}^{\rm SC}$. 
This explains the $T_{\rm c}^{\rm SC}$ enhancement in Fig.~\ref{fig:phase_eta_U4}(a), and 
can provide a new pathway for enhancing superconducting $T_{\rm c}^{\rm SC}$ in terms of Fermi surface distortion.

However, it should be difficult to
achieve purely $d$-wave like distortions for the kernel, and the terms 
having some other symmetries should in general arise even from purely $d$-wave distortions of the Fermi surface. 
We can elaborate this 
by introducing a parameter $g_k$, 
where $g_k$ represents either 
(\romannumeral1) a spontaneous distortion of the electronic states $[\delta g_k = G(k)-G_{\rm undistorted}(k)]$, or
(\romannumeral2) a small $d$-wave modulation of the Hamiltonian $(\delta {\cal H}=\sum_{{\boldsymbol k},\sigma} \delta g_k c_{{\boldsymbol k},\sigma}^{\dag}c_{{\boldsymbol k},\sigma})$.
Then we can expand the interaction kernel in $g_k$, which gives, 
up to the second-order,
\begin{equation}
	K(k,k^{\prime}) \rightarrow K(k,k^{\prime})
	+ \sum_p \frac{\delta K}{\delta g_p} \delta g_p + \frac{1}{2}\sum_{p,q} \frac{\delta^2 K}{\delta g_p \delta g_q} \delta g_p \delta g_q,
\end{equation}
and the effect on the eigenvalue $\lambda$ reads
\begin{align}
	\delta \lambda_{\rm max}^{\rm (2)} &= \sum_{i,k,k^{\prime}} \frac{|\phi^{*}_{\rm max}(k) \sum_p \frac{\delta K}{\delta g_p} \delta g_p \phi_{\rm i}(k^{\prime})|^2 }{\lambda_{\rm max}-\lambda_{\rm i}} \notag \\
			&-\sum_{k,k^{\prime}} \phi^{*}(k) {\frac{1}{2}\sum_{p,q} \frac{\delta^2 K}{\delta g_p \delta g_q} \delta g_p \delta g_q} \phi(k^{\prime}).
\label{second-order}
\end{align}
We can see that whether $T_{\rm c}^{\rm SC}$ can be enhanced depends on the second term on the right-hand side of Eq.~(\ref{second-order}).
From this we expect that the enhancement tends to occur when the
{\it second-largest} 
eigenvalue is close to the largest one, for which the first 
term on the right-hand side of Eq.~(\ref{second-order}) becomes dominant.

\begin{figure}[t]
\begin{centering}
\includegraphics[width=1\columnwidth]{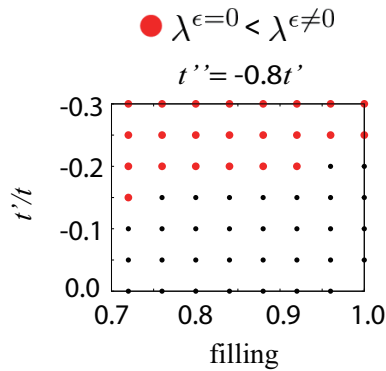}
\par\end{centering}
\caption{(Color online) Comparison of the eigenvalue $\lambda$ calculated with RPA
between the four-fold symmetric Fermi surface $(\epsilon = 0)$ and the distorted Fermi surface $(\epsilon =0.01)$
with red (black) dots representing the case of $\lambda^{\epsilon=0} < \lambda^{\epsilon=0.01}$ ($\lambda^{\epsilon=0} > \lambda^{\epsilon=0.01}$)
for $\beta t =5, U/t=2.0$.  
The horizontal axis corresponds to the band filling, 
while the vertical axis is $t^{\prime}/t$, with $t^{\prime \prime}=-0.8t^{\prime}$
(which includes the parameter set used in Fig.~\ref{fig:tp-dependence}).
}
\label{fig:compare_weak-nematic}
\end{figure}

We can in fact check this argument in the weak-coupling case. 
To obtain qualitative tendencies, we have performed a 
RPA calculation at a relatively high temperature for various values of parameters to 
compare the four-fold symmetric case with the distorted Fermi surface
by making the nearest-neighbor hopping slightly anisotropic, 
 $t_x=1+\epsilon, t_y=1-\epsilon $, by hand 
with the first line in Eq.~(\ref{dispersion}) becoming 
$\epsilon({\boldsymbol k}) = -2t_x{\rm cos}\; k_x -2t_y {\rm cos}\; k_y$. 
In the RPA, we ignore the self-energy effect in the 
Eliashberg Eq.~(\ref{eliashberg}).  

When we compare 
the eigenvalue under a distortion $\lambda^{\epsilon = 0.01}$ with 
$\lambda^{\epsilon = 0}$ for the symmetric case, 
the result in Fig.~\ref{fig:compare_weak-nematic} for $U/t=2.0, \beta t =5$ 
shows that we do have a region (marked with 
red circles representing $\lambda^{\epsilon = 0.01}>\lambda^{\epsilon = 0}$) 
in which the distortion enhances the eigenvalue.  
This effect tends to occur away from half-filling, and 
for larger values of distant-neighbor hopping $t^{\prime}, t^{\prime \prime}$ 
(i.e., more frustrated cases).  
Thus we can confirm that the enhancement of the superconductivity 
by small distortions indeed occurs at least 
in the weak-coupling limit where we can ignore the self-energy effect. 

It has been known that the gap symmetry (for the leading eigenvalue) 
tends to be changed 
for higher doping or more frustrated cases\cite{RPA-wide-parameter-range}.  
The present result suggests that 
the $T_{\rm c}^{\rm SC}$ enhancement arising from the distortion tends to occur
around the boundary for the gap symmetry to change where the
leading and sub-leading eigenvalues are close to each other. 
This is also consistent with the above result for 
the $t^{\prime}$ dependence in FLEX+DMFT (Fig.~\ref{fig:tp-dependence}), where the
enhancement of $T_{\rm c}^{\rm SC}$ occurs for $(t^{\prime}, t^{\prime \prime}) = 
(-0.20, 0.16)$.  
We also notice that the structure of Eq.~(\ref {second-order}) is reminiscent of the pseudo Jahn-Teller effect,
in which a Jahn-Teller-like distortion occurs without degeneracies 
due to the second-order effect of the distortion \cite{SOJT}. 
In this context we can also recall a well-known property 
that, if the eigenvalues are degenerate 
(e.g. for $p+ip$-pairing), 
$T_{\rm c}^{\rm SC}$ can be enhanced by the strain effect \cite{strain-effect-p+ip}. 
From these, the present result may also 
be viewed as a possibility for this kind 
of $T_{\rm c}^{\rm SC}$ enhancement revealed even for the (non-degenerate) $d$-wave regime in the 2D square lattice Hubbard model.

\section{Summary and discussions}
We have employed FLEX+DMFT approach
to study the interplay of Pomeranchuk instability and superconductivity in correlated electron systems.
We have revealed that the superconductivity with the distorted Fermi surface 
has the symmetry of the gap function changed from $d$-wave to $d+s$,
consistent with the previous studies \cite{DCA-nematic,nematic-dSC-DE-GWF}.
We have found that 
the Fermi surface distortion can enhance the superconducting 
$T_{\rm c}^{\rm SC}$ 
in the overdoped regime.  
We have explained this enhancement 
in terms of the perturbation for small distortions, and also 
with RPA in the weak-coupling regime.  
The Pomeranchuk $T_{\rm c}^{\rm PI}$ is found to be 
much more sensitive to Fermi surface warping and 
the position of the van Hove filling 
than the superconducting transition temperature.

In the main parameter set for the present calculation, the Pomeranchuk $T_{\rm c}$-dome 
appears in the overdoped region,
while experimentally 
the electronic nematicity is mainly observed in the underdoped regime.
If the nematicity in the cuprates comes from the Pomeranchuk instability, 
then the present result suggests that it should 
strongly depend on the component materials that can have various 
values of second- ($t'$) and third-neighbor ($t''$) hoppings: 
For instance, ${\rm La}_{2-x}{\rm Sr}_{x}{\rm CuO}_4$ 
with smaller $t', t''$ has 
the van Hove filling sitting around 20\% doping \cite{LSCO-VHS,LSCO-ARPES-2006,
sakakibara}, 
which is close to the situation given in Fig.~\ref{fig:tp-dependence}(c),
where the Pomeranchuk $T_{\rm c}$-dome appears 
around the optimal to underdoped regimes. 

If we comment on the method, 
FLEX+DMFT, despite being an improvement 
over FLEX or DMFT, still overestimates 
the non-local self-energy effect.  
For more accurate estimates, other methods (e.g. diagrammatic expansion in two-particle level as in D$\Gamma$A \cite{DGA}
or dual fermion method \cite{DF}) will be needed.  
Also, we have assumed here translationally invariant systems, 
while the study of the interplay between superconductivity and charge 
instabilities involving finite wave vectors will be another 
interesting future work. 

\section{Acknowledgement}
The present work was supported by JSPS KAKENHI Grant Number JP26247057, and 
by ImPACT Program of Council for Science, Technology and
Innovation, Cabinet Office, Government of Japan (No. 2015-
PM12-05-01) from JST (HA), and by the
Advanced leading graduate course for photon science (MK), 
and by JSPS KAKENHI Grant Number JP25800192 (NT).

\appendix
\section{Effect of the DMFT vertex $\Gamma_{\rm DMFT}$ } \label{appendix}
According to the formulation in Ref.\onlinecite{FLEX+DMFT},
we should consider the local anomalous self-energy $\Delta_{\rm loc}$ coming from the DMFT functional.
Then the linearized Eliashberg equation becomes
\begin{align}
\lambda \Delta(k) = -\frac{1}{N_{\bm{k}}\beta}\sum_{k^{\prime}}
[V_{\rm eff} &(k-k^{\prime})+\Gamma_{\rm DMFT}(\omega_n,\omega_m)] \notag \\
             &\times |G(k^{\prime})|^2 \Delta(k^{\prime}),
\end{align}
where $\Gamma_{\rm DMFT}={\delta \Delta_{\rm loc}}/{\delta F}$ 
is the functional derivative of
the local anomalous self-energy.  
While this term can be ignored for studying pure $d$-wave pairing as in the previous paper \cite{FLEX+DMFT}, 
we examine its effect on the $d+s$ pairing here.
We consider this effect to be small, because the additional term is an 
extended $s$-wave (nonlocal) pairing rather than the ordinary $s$-wave, 
so that a cancellation should occur in the momentum summation.
To check this along the argument in the main text, 
we can calculate the lower bound for the maximal eigenvalue when $\Gamma_{\rm DMFT}$ is considered
without calculating $\Gamma_{\rm DMFT}$ directly.
From the eigenvector $\Delta_{\rm max}$ of Eq.~(\ref{eliashberg}),
we extract the part of the gap function that is not affected by $\Gamma_{\rm DMFT}$ as
\begin{equation}
\Delta^{\prime}(k) = \Delta_{\rm max}(k) - \frac{\sum_{\boldsymbol{k}} |G(k)|^2 \Delta_{\rm max}(k)}{\sum_{\boldsymbol{k}}|G(k)|^2}.
\end{equation}
Then a quantity,
\begin{equation}
\lambda^{\prime} = -\frac{\sum_{k,k^{\prime}} \Delta^{\prime *}(k)|G(k)|^2 V_{\rm eff}(k-k^{\prime})
|G(k^{\prime})|^2 \Delta^{\prime}(k^{\prime})}
{\sum_k \Delta^{\prime *}(k)|G(k)|^2 \Delta(k)},
\end{equation}
gives the lower bound for the maximal eigenvalue when $\Gamma_{\rm DMFT}$ 
is considered.  We have actually confirmed 
that the difference between $\lambda^{\prime}$ and $\lambda$ (without $\Gamma_{\rm DMFT}$) is very small, $(\lambda-\lambda^{\prime})/\lambda < 0.01$.
Thus we can conclude the effect of the DMFT vertex $\Gamma_{\rm DMFT}$
does not significantly change the the result for the $T_{\rm c}^{\rm SC}$ enhancement.

\newcommand{\PR}[3]{Phys. Rev. \textbf{#1}, #2 (#3)}
\newcommand{\PRL}[3]{Phys. Rev. Lett. \textbf{#1}, #2 (#3)}
\newcommand{\PRA}[3]{Phys. Rev. A \textbf{#1}, #2 (#3)}
\newcommand{\PRB}[3]{Phys. Rev. B \textbf{#1}, #2 (#3)}
\newcommand{\JPSJ}[3]{J. Phys. Soc. Jpn. \textbf{#1}, #2 (#3)}
\newcommand{\arxiv}[1]{arXiv:#1}
\newcommand{\RMP}[3]{Rev. Mod. Phys. \textbf{#1}, #2 (#3)}

\end{document}